\documentclass[runningheads]{svmult}

\usepackage{makeidx}   
\usepackage{graphicx}  
\usepackage{subeqnar}  
\usepackage{multicol}  
\usepackage{physmubb}  
\makeindex             

\begin{document}
\title*{Jet Production at CDF}
\author{Mario Mart\'\i nez\\
{\em Fermi National Accelerator Laboratory}\\
{(on behalf of the CDF Collaboration)}}
%
%
%
%

\maketitle              

\section{Introduction}

The Run 2 at Tevatron will define a new 
level of precision for QCD studies in hadron collisions. 
Both collider experiments, CDF and D0, expect
to collect up to $15 \  {\rm fb}^{-1}$ of data in this new 
run period. The increase in instantaneous luminosity, 
center-of-mass energy (from 1.8 TeV to 2 TeV) and the improved 
acceptance of the detectors will allow stringent tests of the 
Standard Model (SM) predictions in extended regions of jet transverse 
energy, $E^{\rm jet}_T$, and jet pseudorapidity, $\eta^{\rm jet}$. 

In the following, a review of some of the most important QCD results from 
Run 1  is presented, together with  first preliminary Run 2 
measurements (based on the very first data collected by the experiment)  
and future prospects as the integrated luminosity increases.

\section{Inclusive Jet Production}

\begin{figure}
\centerline{
\includegraphics[width=0.5\textwidth]{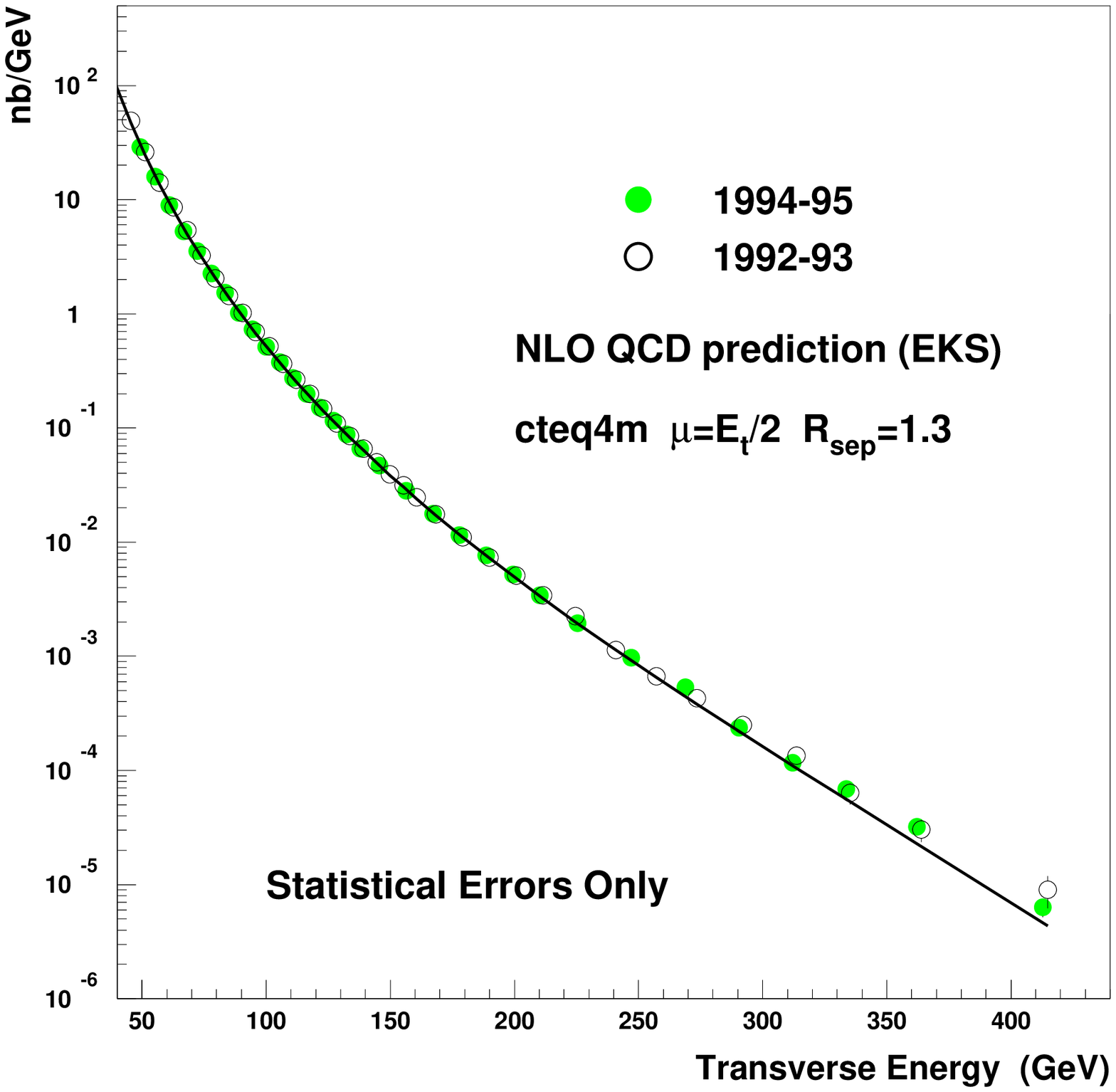}
\includegraphics[width=0.5\textwidth]{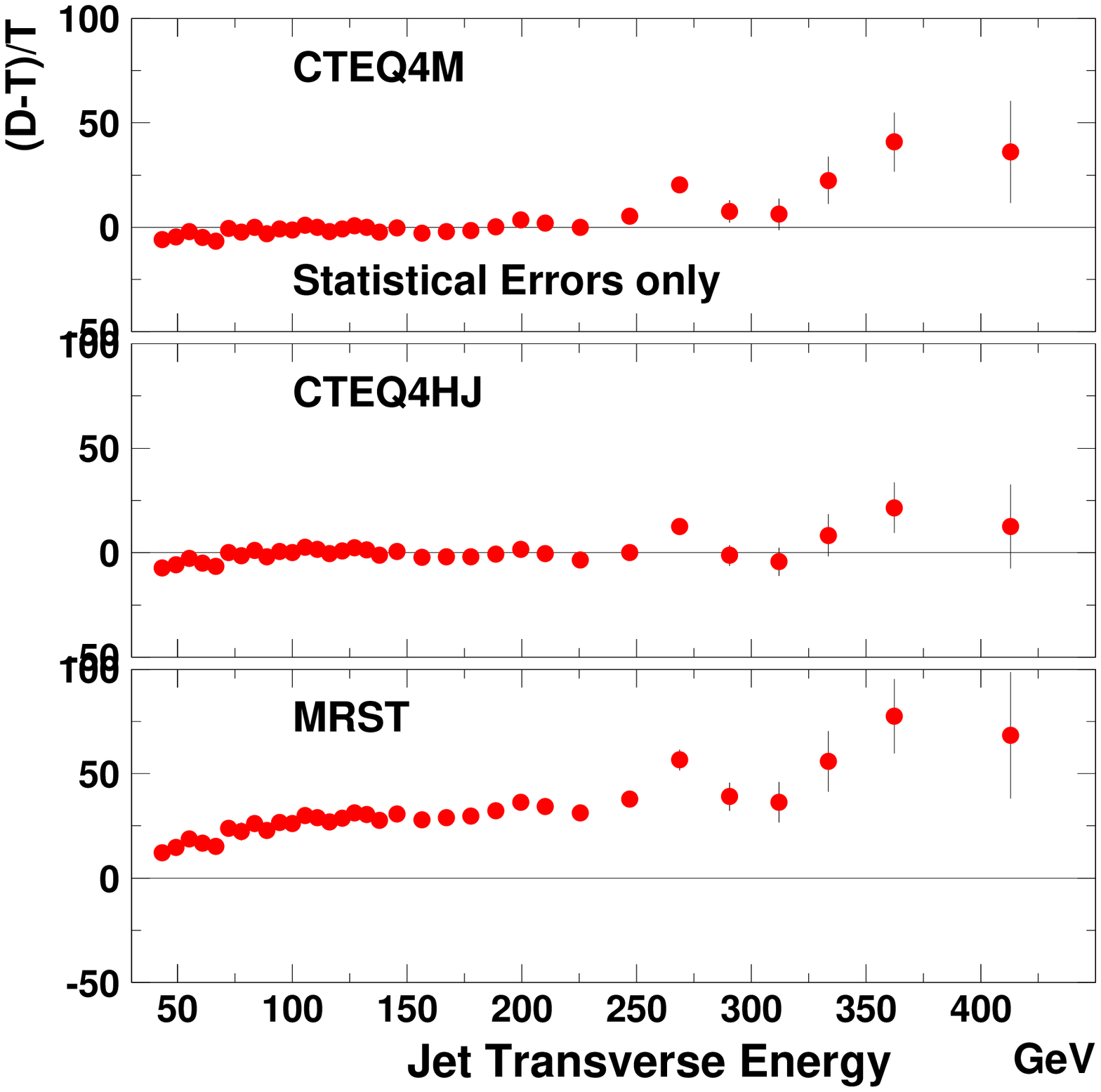}
}
\caption[]{(Left) Measured inclusive jet cross section as a function 
of $E^{\rm jet}_T$ for jets in the region 
$0.1< | \eta^{\rm jet}| < 0.7$. The measurement is compared to NLO QCD
predictions.(Right)  Ratio (Data - Theory)/Theory for the measured jet cross
section where different parton density functions are considered in 
the theoretical predictions.}
\label{fig1}
\end{figure}

The CDF Run 1 inclusive jet cross section measurements 
\cite{jetincl}, performed for  jets in the region $0.1 < | \eta^{\rm jet}| < 0.7$ 
and $E^{\rm jet}_T > 50$ GeV, 
showed an excess with respect to the NLO calculation in the region 
$E^{\rm jet}_T > 300$ GeV which initially suggested a possible 
signal for new physics (see Fig.~\ref{fig1}-left). However, a detailed 
revision  
of theoretical uncertainties in the NLO calculations indicates 
that the SM predictions at high-$E^{\rm jet}_T$ suffer from 
large uncertainties mainly due to the little knowledge of the  
proton's gluon distribution at high-$x$. The CTEQ 
Collaboration showed that it is possible to describe the CDF measurements
(see Fig.~\ref{fig1}-right) by increasing the amount of gluons 
in the proton at high-$x$ ($x > 0.3$) 
without  affecting the good description of 
the rest of the data used in the global fits like, for example, the 
very precise DIS data. Nowadays, the Tevatron high-$E^{\rm jet}_T$ data, 
although still  has very little statistical power,  is
being used, together with prompt-photon data  
from fixed target experiments, to constrain the gluon 
distribution at high-$x$.

The new Run 2 data will allow better and  more precise jet 
measurements. The increase in the center-of-mass energy will  
extend the measured cross sections from $E^{\rm jet}_T \sim 450$ GeV to 
$E^{\rm jet}_T \sim 600$ GeV, and re-explore possible 
deviations from the SM predictions. 
Independent cross section measurements for forward-forward 
and central-forward dijet production will be essential to constrain the 
gluon distribution at high-$x$ and separate an eventual signal 
for new physics from the current SM uncertainties. Forward jet measurements
are not expected to have any contribution from new physics due to the 
fact that the maximum reachable $E^{\rm jet}_T$ is limited 
to $E^{\rm jet}_T \sim 200$
GeV, but have a sensitivity to the proton's gluon distribution
 similar to that of the central jet measurements. 

The CDF Run 1 analyses used the cone algorithm \cite{cone} to search for 
jets, define jet observables and measure jet cross sections. During 
the past few years different theoretical problems of the cone algorithm
were pointed out, namely: the infrared and collinear 
sensitivity of the defined
cross sections and the difficulty to translate
 the experimental prescription for merging jets to an equivalent 
procedure at the parton level in theoretical calculations.   
The longitudinally invariant $K_T$ algorithm \cite{kt}, initially 
used in $e^+e^-$ collisions, was proposed for $ep$ and  $pp$($p\overline{p}$) 
experiments \cite{soper}. It has been already used by the DIS experiments
at HERA \cite{oscar} with so great success that the cone algorithm has been 
abandoned. Therefore, one of the main goal of the QCD program at Tevatron
in Run 2 will be the study of the performance of  
$K_T$ algorithms in a hadron-hadron environment.        

\section{Three-jet Production}

CDF recently presented a measurement of inclusive three-jet production 
compared to NLO calculations based on Run 1 data. Jets 
were searched for using a cone algorithm with radius R=0.7. The events 
were required to have at least three jets with $E^{\rm jet}_T > 20$ GeV and
$| \eta^{\rm jet}| < 2.0$.  In order to compare with NLO 
calculations, additional cuts were applied. The sum of the transverse energy
of the jets was required to be above 320 GeV and a cut on the minimum 
separation between jets of 1.0 unit ($\eta - \phi$ space) was used.  

\begin{figure}[h]
\centerline{
\includegraphics[width=0.45\textwidth]{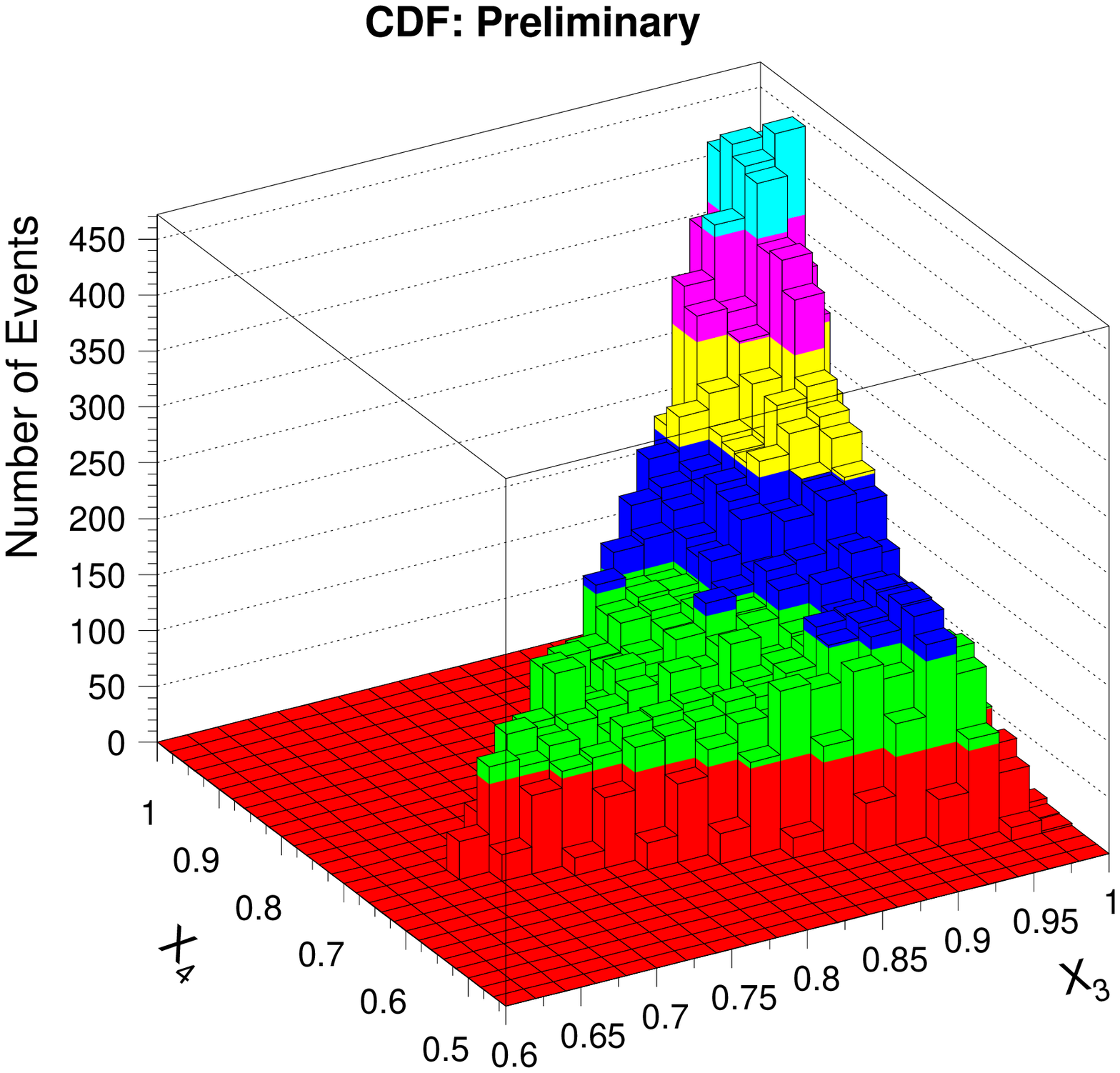}
\includegraphics[width=0.5\textwidth]{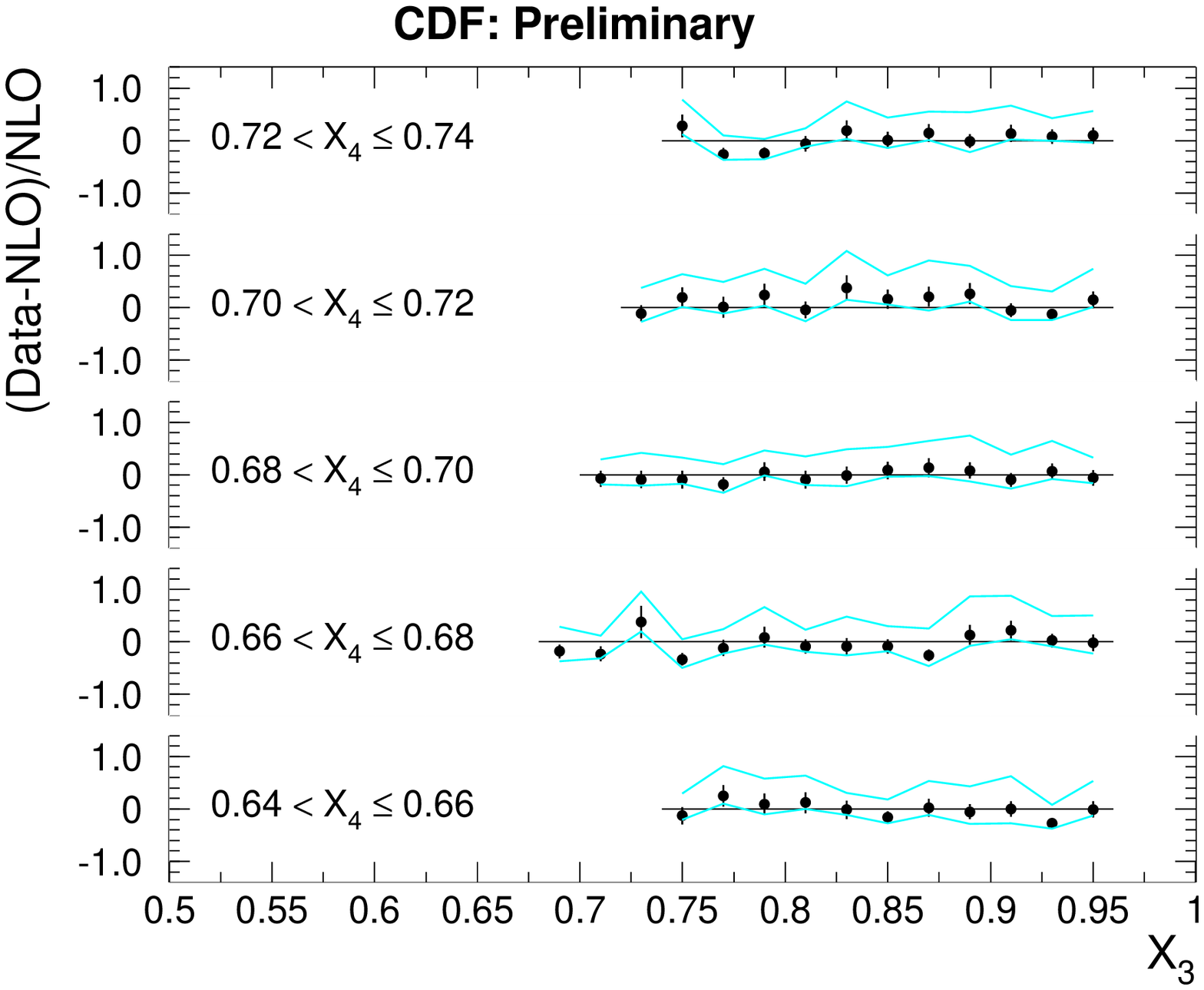}
}
\caption[]{(Left) Distribution of the measured three-jet events in 
the $X_3-X_4$ plane. (Right) Ratio (Data - NLO)/NLO for the differential
cross section as a function of $X_3$ measured 
in the region  $0.64 \leq X_4 \leq 0.74$.}
\label{fig2}
\end{figure}

The topology of the three-jet final state was studied using Dalitz 
variables in the center-of-mass of the three-jet system:

\begin{equation}
X_i = \frac{2 \cdot E^{\rm jet}_i}{M_{\rm 3 jets}} , \ \ i = 3,4,5 \label{eq1}
\end{equation}
where $M_{\rm 3 jets}$ denotes the invariant mass of the three jets and
the  jets are sorted in energy in such a way that $X_3 > X_4 > X_5$, where 
$X_3 + X_4 + X_5 \equiv 2$. Figure ~\ref{fig2}-left shows the distribution
of the measured three-jet events in the ($X_3-X_4$) plane. Different event 
topologies are observed, including {\it{Mercedes-Benz Star}} type of  
events with $X_3 \sim 0.7$ and  $X_4 \sim 0.7$. However, the topologies
are dominated by those configurations with a soft third jet where $X_{3,4} \sim 0.95$. 

The differential cross section as a function of $X_3$, measured in 
different regions of $X_4$, was compared to  NLO calculations 
\cite{3jetnlo}. 
Figure \ref{fig2}-right shows the comparison between data and 
theory in the region $0.64 \leq X_4 \leq 0.74$. Reasonable 
agreement was observed in the whole ($X_3 -X_4$) plane. The total measured 
three-jet production cross section, integrated over the Dalitz plane, 
was $\sigma^{\rm 3 jets} = 466 \pm 2 ({\rm stat.}) {}^{+206 }_{-71} ({\rm syst.})$ pb, consistent with the NLO prediction $\sigma^{\rm 3 jets}_{\rm NLO} = 402 \pm 3$ pb.

\section{Study of Jet Shapes in Run 2}

The first 16~${\rm pb}^{-1}$ of dijet data collected by CDF in Run 2 
were used to measure the jet shapes for jets in the range 
$30 < E^{\rm jet}_T < 135$ GeV and $| \eta^{\rm jet} | < 2.3$. Jets 
were searched for using a cone algorithm with R=0.7 starting from the 
energy deposits in the calorimeter towers, and the jet variables  
were reconstructed according to the Snowmass convection.
The integrated jet shape, $\Psi(r)$, is defined as the average 
fraction of the transverse energy of jet with lies inside an cone of 
radius $r$ concentric to the jet cone:
\begin{equation}
\Psi(r) = \frac{1}{N_{\rm jets}} \frac{\sum E_T (0,r)}{E^{\rm jet}_T}, \ \ \Psi(r=R) = 1, \label{eq2}
\end{equation}
where the sum runs over the calorimeter towers belonging  
to the jet.

\vspace{-1 cm}
\begin{figure}
\centerline{
\includegraphics[width=0.85\textwidth]{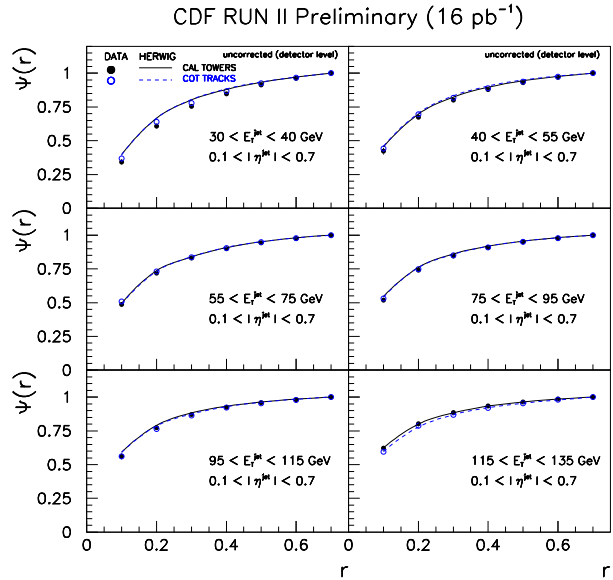}
}
\caption[]{Measured  integrated jet shape, $\Psi(r)$, computed 
using both calorimeter towers (back dots) and tracks 
(open circles)  for jets in the 
region $0.1 < | \eta^{\rm jet} | < 0.7$ and $30 < E^{\rm jet}_T < 135$ GeV. The
measurements are compared to HERWIG predictions including CDF detector 
simulation.}
\label{fig3}
\end{figure}

 In addition,  for jet with $0.1 < | \eta^{\rm jet} | < 0.7$, the jet shapes 
were measured using tracks and  a similar 
expression as Eq.~\ref{eq2} where $E^{\rm jet}_T$ was substituted by the 
scalar sum of the tracks inside the cone of the jet. 
Figure~\ref{fig3} shows 
the measured jet shapes using both calorimeter towers and tracks compared to 
the predictions from HERWIG MC \cite{herwig}. The measurements performed using
calorimeter and tracking are in excellent agreement. For a given 
fixed distance $r_0$,  the measured $\Psi(r=r_0)$ increases with 
$E^{\rm jet}_T$ indicating that
the jets become narrower. The measured jets shapes are well described by
the HERWIG MC predictions.
Similar studies with b-tagged jets
will be necessary to test our knowledge of b-quark jet fragmentation processes in 
hadronic interactions, which is essential for future precise Top and Higgs 
measurements.

\section{Study of the Underlying Event}

The hadronic final states from QCD processes in $p \overline{p}$ collisions at 
Tevatron are characterized by the presence of soft underlying 
emissions, usually denoted as {\it{underlying
 event}}, in addition to highly energetic jets coming from the 
hard interaction. The underlying event contains contributions from 
initial- and final-state soft gluon radiation, secondary semi-hard partonic
interactions and interactions between the proton and anti-proton remnants that
cannot be described by perturbation theory. These processes  must be 
approximately modeled using  MC programs tuned to describe the data.

\begin{figure}
\centerline{
\includegraphics[width=0.3\textwidth]{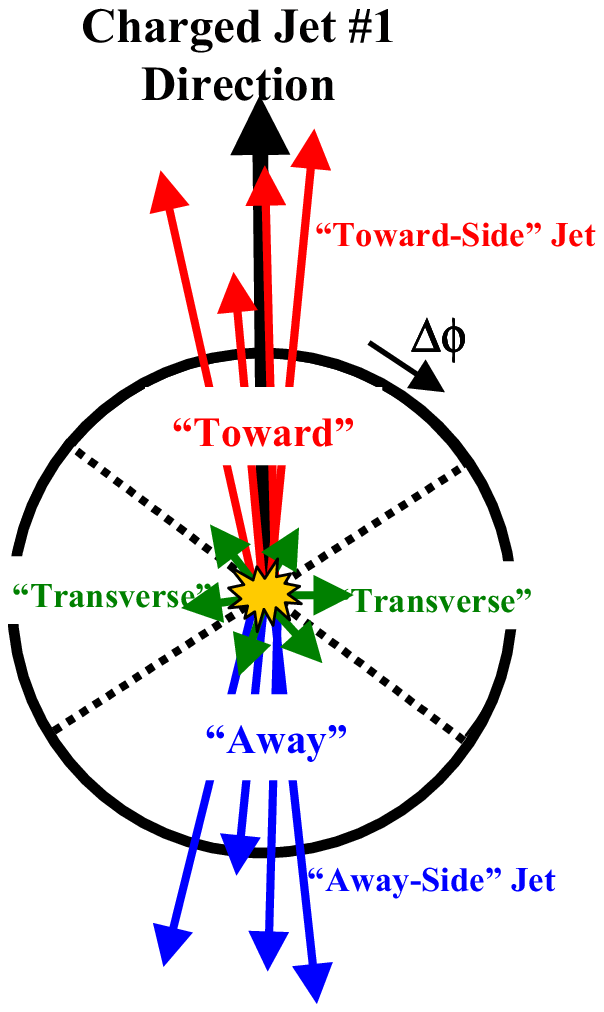}
\includegraphics[width=0.7\textwidth]{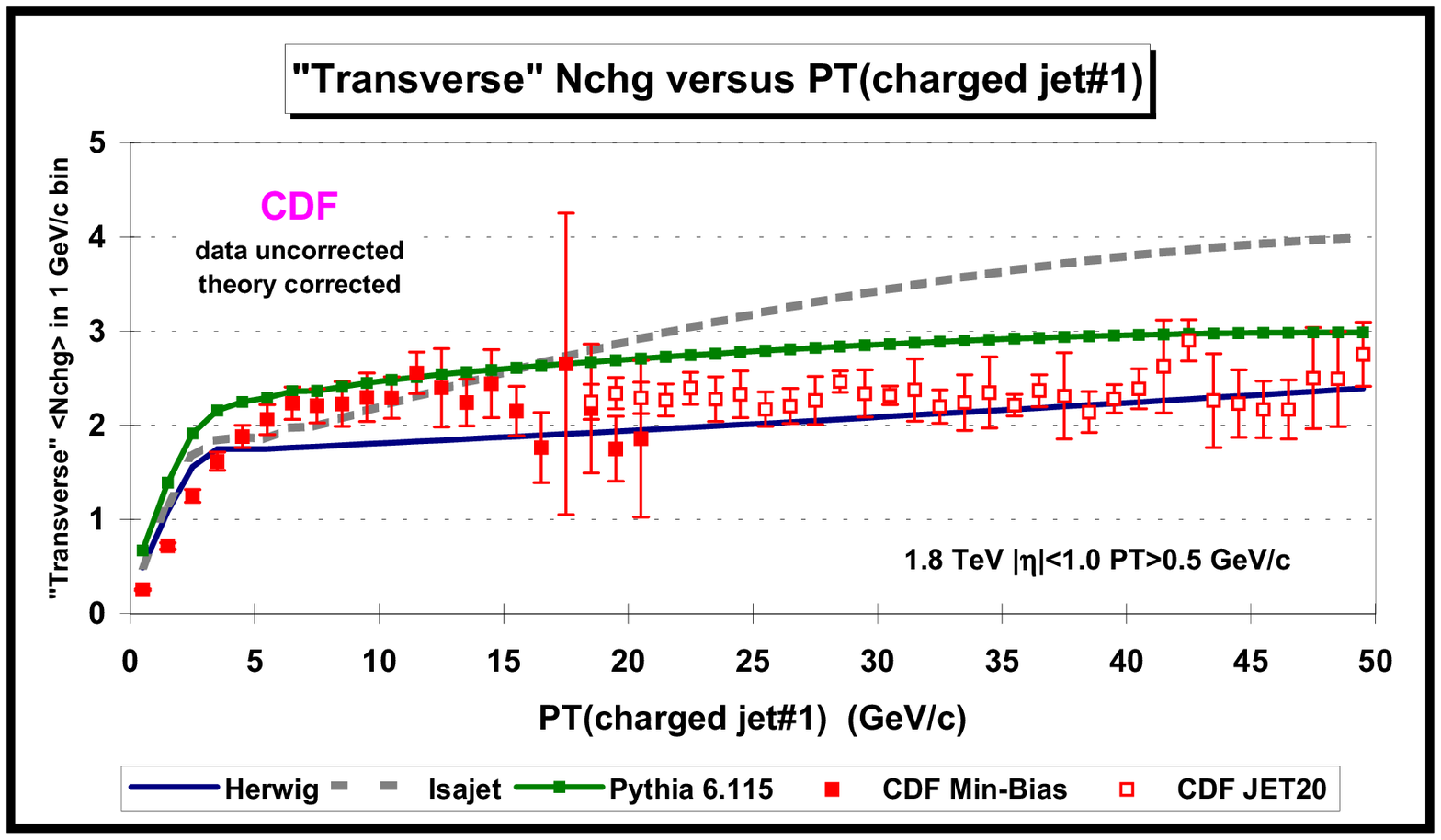}
}
\caption[]{(Left) Scheme of the different $\phi$ regions defined 
around the leading jet. (Right) Measured average track multiplicity 
in the transverse region as a function of the $P_T^{\rm jet}$
of the leading jet. The measurements are compared to different MC models.}
\label{fig4}
\end{figure}

 The jet energies measured in the detector contain an underlying event 
contribution that has to be subtracted in order to compare the measurements 
to pQCD predictions.
Therefore, a proper understanding of this underlying event contribution 
is crucial to reach the desired precision in 
the measured jet cross sections.
In the analysis  presented here \cite{under}, the underlying 
event  in dijet
events was studied by looking at regions well separated from the leading 
jets, where its contribution is expected to dominate the
observed hadronic activity. Jets were reconstructed using  tracks with 
$p_T^{\rm track} > 0.5$ GeV and $|\eta^{\rm track}|<1$ and a cone
 algorithm (Snowmass convection) with R=0.7 in the ($\eta-\phi$) space.  
Jets were sorted in $P_T^{\rm jet}$ and the leading jet defined the 
direction $\phi = 0$. The $\phi$ space around the leading jet was divided
in three regions: {\it{towards}}, {\it{away}} 
and {\it{transverse}} (see Fig.~\ref{fig3}-left), and the transverse
region was assumed to reflect the underlying event contribution. 
Figure~\ref{fig4}-right shows the average track multiplicity in the 
transverse region as a function of $P_T^{\rm jet}$ of the leading jet.
The observed plateau indicates that the underlying event activity 
is, to a large extend, independent from the hard interaction.
Figure~\ref{fig4}-right shows the comparison with 
different MC predictions with default parameters. It becomes 
clear that the measured track multiplicities provide the necessary input to 
tune the  different parameters of the underlying-event models.

\section{Study of W+$N_{\rm jet}$ Production}

   A detailed study of hard processes involving the associated 
production of a W boson and a given number of jets in the final state
 is a main goal of the CDF physics program in Run 2. These processes 
constitute the biggest background to Top and Higgs production in 
hadron colliders.
Therefore, precise measurements of W+$N_{\rm jets}$ cross sections will 
be essential to test the NLO QCD calculations used in order to estimate
QCD-related backgrounds to Top/Higgs signals.

\begin{figure}
\centerline{
\includegraphics[width=0.5\textwidth]{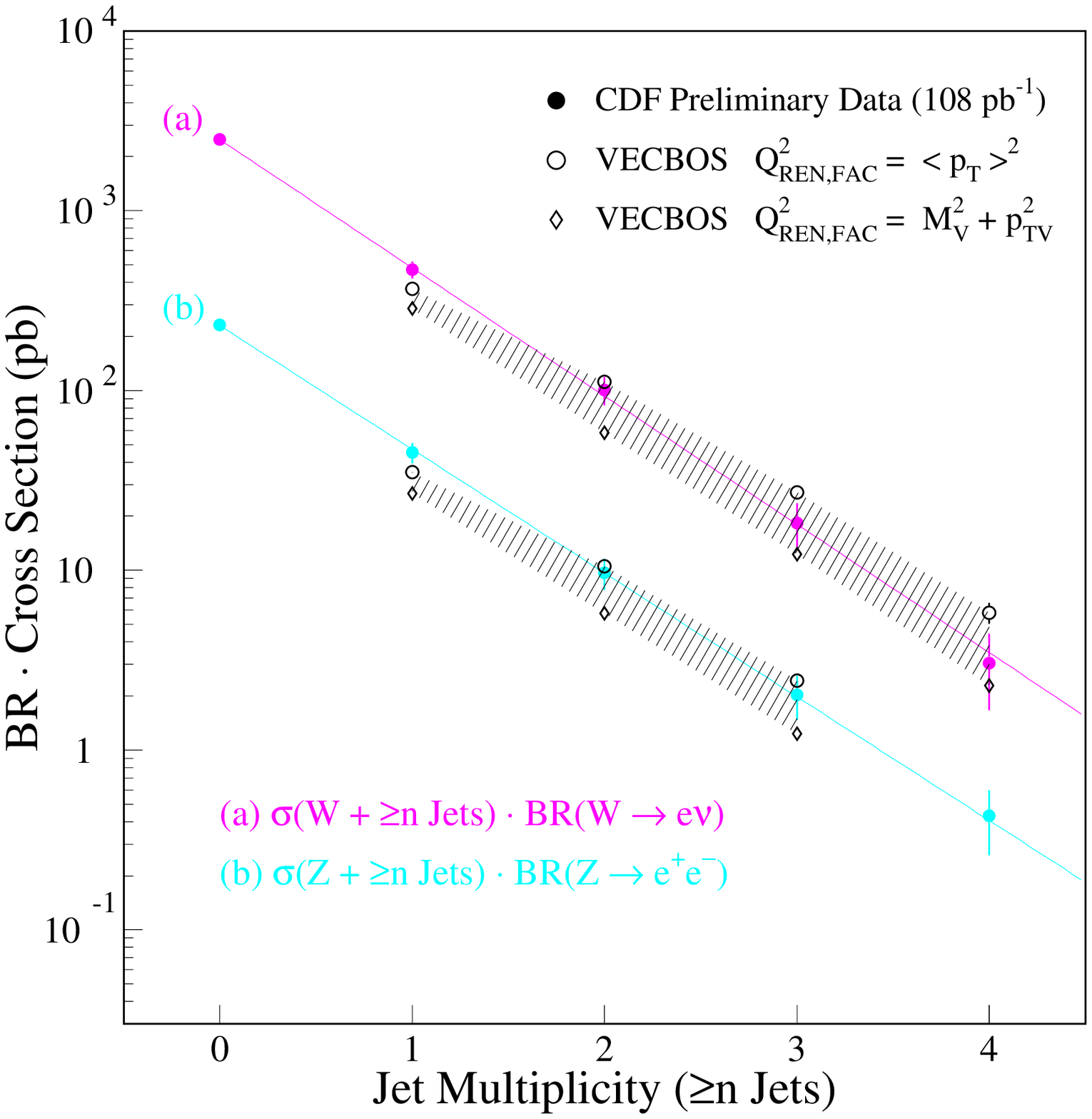}
\includegraphics[width=0.5\textwidth]{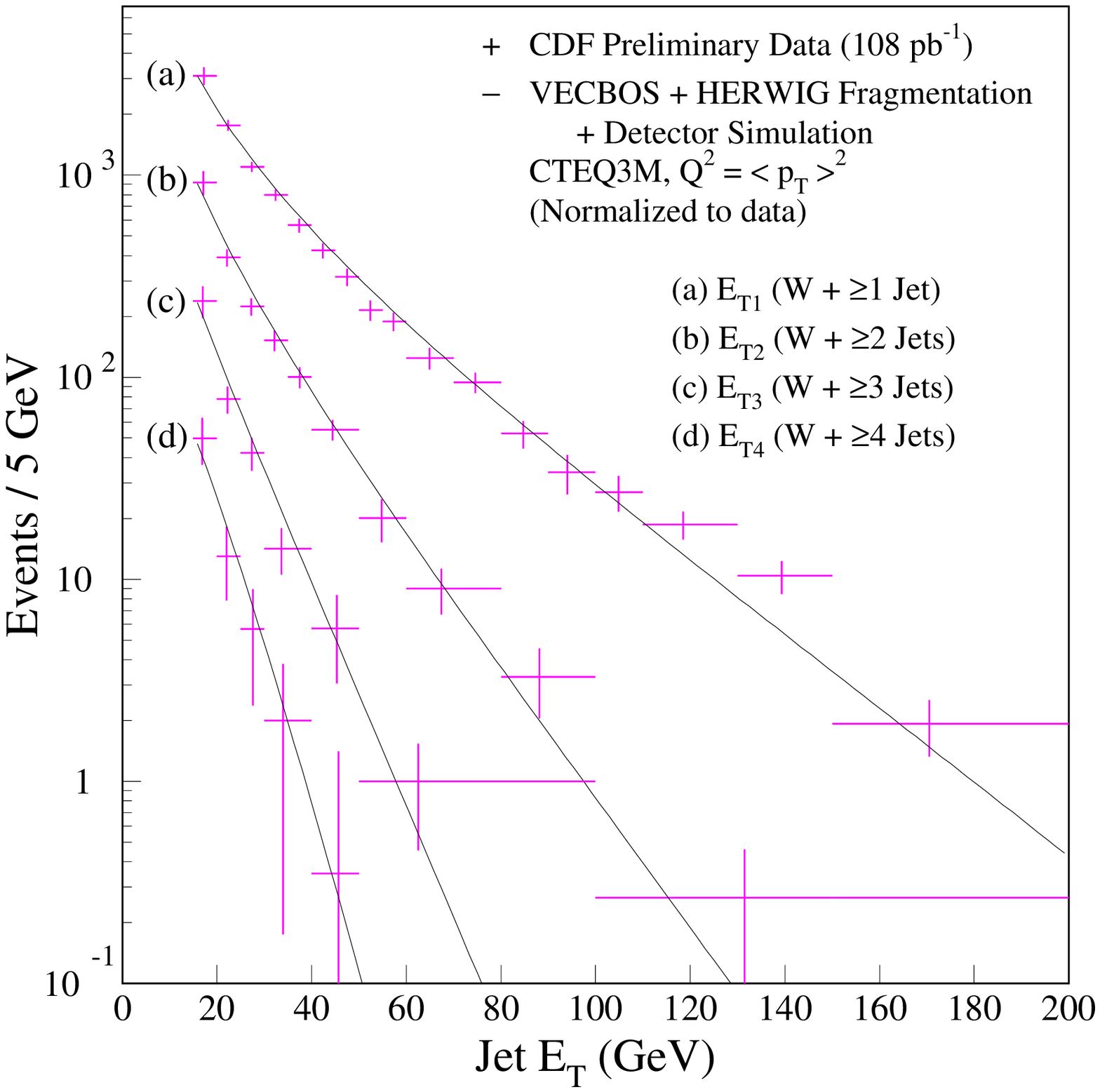}
}
\caption[]{(Left) Measured W+$N_{\rm jet}$ (and Z+$N_{\rm jet}$)
 cross sections as a function of 
jet multiplicity compared to VECBOS predictions. (Right) $E_T^{\rm jet}$ 
spectrum of the less energetic jet in $W+N_{\rm jet}$ production compared
to VECBOS+HERWIG predictions.}
\label{fig5}
\end{figure}

In Run 1, CDF measured the cross section for W+$N_{\rm jet}$ 
production with $N_{\rm jet} \leq 4$ \cite{wjets}. The measurements 
were compared to an {\it{enhanced 
leading-order prediction}} based on leading-order calculations 
(as implemented in VECBOS \cite{vecbos}) interfaced 
to HERWIG for additional gluon radiation and 
hadronization, see Fig.~\ref{fig5}-left.
The measured cross sections were described by the calculations 
that, however, show a large renormalization scale uncertainty. The 
$E_T^{\rm jet}$ distribution for the less energetic jet in  
W+$N_{\rm jet}$ events (see Fig.~\ref{fig5}-right) which is 
sensitive to the additional gluon 
radiation from HERWIG, was also well described by the enhanced 
leading-order prediction.

During the last years a number of new Boson+$N_{\rm jet}$ programs have 
become available \cite{mcs} which include larger jet  multiplicities in 
the final state, in addition to NLO calculations for the W+dijet case. These 
different programs are being interfaced to 
parton-shower models  and will make possible precise comparison with Tevatron 
data as the recorded luminosity of the experiments increases. These measurements will 
become the testing ground  for a future 
discovery of new particles either at Tevatron or  at the LHC.

%

\end{document}